\documentclass[twocolumn,aps,prl,epsf,psfig,showpacs]{revtex4}
\input{epsf}
\input epsf.sty
\input psfig.sty

\begin{document}

\title{Dynamic response of trapped ultracold bosons on optical
lattices}

\author{G.G. Batrouni$^1$, F.F.~Assaad$^2$, R.T.~Scalettar$^3$,
  P.J.H. Denteneer$^4$} 
\affiliation{$^1$Institut Non-Lin\'eaire de Nice, Universit\'e de
  Nice--Sophia Antipolis, 1361 route des Lucioles, 06560 Valbonne,
  France} 
\affiliation{$^2$Institut f\"ur Theoretische Physik und Astrophysik,
Universit\"at W\"urzburg, Am Hubland, D-97074 W\"urzburg, Germany} 
\affiliation{$^3$Physics Department, University of California, Davis, CA
  95616, USA} 
\affiliation{$^4$Lorentz Institute, Leiden University, P. O. Box 9506,
2300 RA  Leiden, The Netherlands}

\begin{abstract}
We study the dynamic response of ultracold bosons trapped in
one-dimensional optical lattices using Quantum Monte Carlo simulations
of the boson Hubbard model with a confining potential.  The dynamic
structure factor reveals the inhomogeneous nature of the low
temperature state, which contains coexisting Mott insulator and
superfluid regions.  We present new evidence for local quantum
criticality and shed new light on the experimental excitation spectrum
of $^{87}$Rb atoms confined in one dimension.
\end{abstract}

\pacs{03.75Hh,03.75.Lm,05.30.Jp}

\maketitle

The advent of Bose--Einstein condensation in cold atomic gases opened
new fields of research merging atomic physics and quantum optics with
condensed matter physics. Scattering experiments common in solid state
physics and liquid helium have been adapted to extract information on
excitation dynamics of atomic condensates resulting in the technique
of {\em Bragg spectroscopy} \cite{Braggspec}.  Initial experiments on
Bose-condensed gases could be well described by mean-field
Gross-Pitaevskii approaches, valid for weak interactions.  However,
recent experiments have reached the regime of strongly interacting
atoms, often by confining the atoms in {\em optical lattices},
leading, for instance, to the destruction of a phase-coherent
superfluid (SF) phase and the establishment of Mott insulator (MI)
phase with squeezed number fluctuations
\cite{NatGreiner,StofEss04,NatParedes}.  The latter state is
considered as promising for use as a large quantum register for
quantum information purposes \cite{JakschQI}.

Ultracold atoms in optical lattices are an almost ideal realisation of
the {\em boson Hubbard} model \cite{Jaksch98}, that has previously
been studied intensely (with different
motivation)~\cite{FisherM,GGBRTS90}.  Much of this previous work
focussed on the equilibrium phase diagram or on the conductivity in
the presence of disorder.  However, as in other condensed matter
systems, a rich source of information on the excitation dynamics is
the {\em dynamic structure factor}, the double Fourier transform of
density-density correlations, which can be measured using Bragg
spectroscopy \cite{Braggspec,DSFBH,roth}.  It is therefore of great
current interest to study the dynamic structure factor of the boson
Hubbard model.  Because of loss of translational invariance due to the
presence of the confining trap, results for the uniform system cannot
be directly applied to the confined system.  This affects both the
static response, the system no longer has a globally incompressible
phase\cite{PRL02}, and the dynamics, since the wave vector is no
longer a good quantum number.

In this paper, we use Quantum Monte Carlo to study the ground state of
the one-dimensional boson-Hubbard model in a confining harmonic
potential:
\begin{eqnarray}
\nonumber
H&=&-t\sum_{i} \left(a^{\dagger}_i a_{i+1} + a^{\dagger}_{i+1} 
 a_i\right) + \\
& & V_T\sum_i (x_i-\frac{L}{2})^2 \, n_i + U\sum_i
 n_i(n_i-1).
\label{hubham}
\end{eqnarray}
where $L$ is the number of sites and $x_i$ is the coordinate of the
$i$th site. The hopping parameter, $t$, sets the energy scale,
$n_i=a^\dagger_i a_i$ is the number operator,
$[a_i,a^\dagger_j]=\delta_{ij}$ are bosonic creation and destruction
operators. $V_T$ sets the confining trap curvature, while the contact
interaction is given by $U$. We use the World Line algorithm for our
simulations. The static properties and state diagram of this model were
studied in \cite{PRL02} where it was shown, among other things, that
the presence of the trap and the destruction of translation invariance
eliminate the quantum phase transitions between the SF and MI phases
and replace it by a co-existence of phases.  Our main results are: (i)
due to co-existence of MI and SF regions in the confined system, the
dynamic structure factor exhibits two excitation branches, one gapped
and one with a linear, phonon-like, dispersion; (ii) these branches
are shown clearly to be connected to the spatial regions containing
the MI and SF phases and (iii) by using an appropriate measure of
local excitations, we provide new evidence for local quantum
criticality.

In the ground state ($T=0$) the dynamic structure factor is given by
\begin{eqnarray}
\label{skw1}
\nonumber
S(k,\omega) &=& \frac{1}{L^2 L_{t}}\sum_t \sum_{r,r^\prime} {\rm
  e}^{-i\omega t} {\rm e}^{ik(r-r^\prime)}\\
& & \times \langle n(r,0) n(r^\prime,t)
  \rangle\\
&=& \sum_m \delta(\omega - E_m + E_0) |\langle 0| n(k,0) | m \rangle
  |^2,
\label{skw2}
\end{eqnarray}
where $L_{t}$ is the number of time slices, $E_m$ the energy of state
$|m\rangle$, and $\langle n(r,0) n(r^\prime,t) \rangle$ the space and
time separated density-density correlation function. With our
normalization we have $ \int {\rm d}\omega S(k,\omega)=N_b S(k)$ where
$ S(k)$ is the static structure factor and $N_b$ the number of
bosons. Equation~(\ref{skw2}) shows that, for a given $k$, the peaks
in $S(k,\omega)$ correspond to the excitation energies. We have
performed all our simulations at $\beta=T^{-1}=10$ and verified that
it is sufficient to study the ground state.  In the World Line
algorithm, it is a simple matter to measure $\langle n(r,0)
n(r^\prime,\tau) \rangle$ where $\tau$ is the imaginary time
separation. The dynamic structure factor is then given by
Eq.~(\ref{skw1}), with $t$ replaced by $-i\tau$ which then requires
the performance of a Laplace, instead of a Fourier, transform. This is
done with the help of the stochastic method (SM)~\cite{stochastic}.

\begin{figure}
\psfig{file=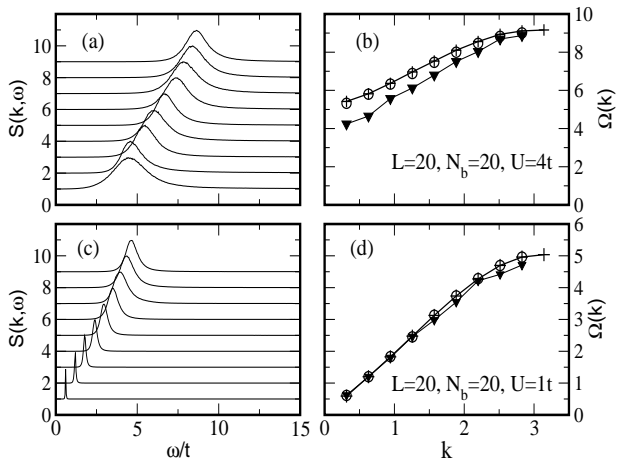,height=3in,width=3.5in,angle=-90}
\caption{(a,c) $S(k,\omega)$ vs $\omega$ and (b,d) $\Omega(k)$ vs $k$
  for $k=2\pi n/L$ with $1\leq n\leq 9$. The system has commensurate
  filling: $L=20, N_b=20, \beta=10$ with $U=4t$ (a,b) and $U=1t$
  (c,d). }
\label{fullfillomega}
\end{figure}

Figure~\ref{fullfillomega} shows $S(k,\omega)$ for the uniform system
in the MI (a) and SF (c) phases. The critical value in one dimension
is $U_{crit}=2.33t$~\cite{monien}. For better visibility, our
$S(k,\omega)$ curves in Figs.~\ref{fullfillomega},\ref{SkwNb25-64}
have been normalized $\sum_{\omega} S(k,\omega)=1$ and do not show the
true spectral weights. For the SF, it is clear that the excitation
energies for small $k$ are linear in $k$, {\it i.e.}  phononic,
leading to a stable SF phase (according to the Landau criterion). On
the other hand, the MI case clearly shows a gap at $\omega\approx 3t$,
where the spectral weight in Fig.~\ref{fullfillomega}(a) starts to be
appreciable. This is in excellent agreement with the value of the gap
of $3t$ measured in~\cite{GGBRTS90} from the static
compressibility. Note that the first peak is at $\omega \approx 4.5t$,
which is larger than the gap, and that the peaks for the MI are rather
wide. This is due to the presence of several closely spaced peaks
which we cannot resolve. Such closely spaced peaks are seen in the
exact diagonalization of smaller systems~\cite{roth} where, for the
MI, the first in the lowest group of peaks does indeed correspond to
the gap.

One can study the excitation spectrum using the dispersion relation
$\Omega(k)$,
\begin{equation}
\Omega(k) = \frac{\int {\rm d}\omega\,\,\omega \,
 S(k,\omega)}{\int {\rm d}\omega\,\,  S(k,\omega)},
\label{dyndisp}
\end{equation}
which can be expressed in terms of purely static quantities using the
$f$-sum rule 
\begin{equation}
\int_{-\infty}^{+\infty} {\rm d}\omega\,\, \omega \, S(k,\omega) = N_b E_k,
\label{fsum}
\end{equation}
\begin{equation}
 E_k = \frac{-t}{L}\left ( {\rm cos}(2\pi k/L)-1\right )
\langle 0| \sum_{i=1}^L
 \left ( a^{\dagger}_i a_{i+1} + a^{\dagger}_{i+1} a_i\right)|0
 \rangle.
\label{ek}
\end{equation}
The dispersion relation is then given by the Feynman result,
\begin{equation}
\Omega(k) = \frac{E_k}{S(k)}.
\label{disp}
\end{equation}
Figure~\ref{fullfillomega} shows $\Omega(k)$ obtained using the
dynamic structure factor, Eq.~(\ref{dyndisp}) (circles), and also
using only static quantities, Eqs.~(\ref{ek}) and ~(\ref{disp})
(pluses) in the MI (b) and SF (d) phases. These two calculations agree
very well and give for the SF phase linear dispersion for small $k$
and a gap for the MI. While $\Omega(k)$ shows clearly the presence of
a gap for MI, it does not give its value since it is an integral over
all peaks that may be present. In addition, we show the positions of
the peaks in both cases (down triangles) which agree extremely well
with $\Omega(k)$ when the peaks are very sharp. This offers a
consistency and precision check on our methods. Our results for
$\Omega(k)$ agree with the small lattice exact diagonalization
results~\cite{roth}.

\begin{figure}
\psfig{file=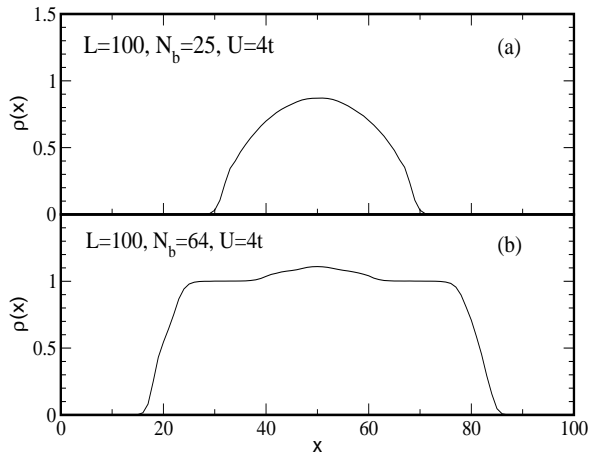,height=3in,width=3.5in,angle=-90}
\caption{Local density profile, $\rho(x)$, for the confined system
  with $L=100$, $V_T=0.008$, $U=4t$: (a) $N_b=25$ (pure SF) and (b)
  $N_b=64$ (co-existence SF-MI).}
\label{profiles}
\end{figure}

\begin{figure}
\psfig{file=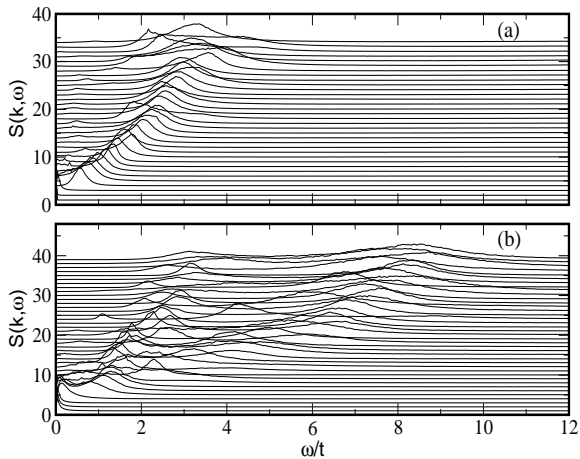,height=3in,width=3.5in,angle=-90}
\caption{$S(k,\omega)$ vs $\omega$ ($k=2\pi n/L$) for the system in
  Fig.~\ref{profiles}, $N_b=25$ (a) and $N_b=64$ (b). The two ridges in
  (b) reflect the SF-MI co-existence, Fig.~\ref{profiles}(b).}
\label{SkwNb25-64}
\end{figure}

We now turn to the confined system where we always take a system size
$L=100$ and $V_T=0.008$. These values and the fillings we use are
comparable to the experimental ones.  Figure~\ref{profiles} shows the
local density profiles for a system in a pure SF phase, (a), and with
coexistence of SF and MI, (b). Figure~\ref{SkwNb25-64} shows the
corresponding $S(k,\omega)$. For the pure SF case,
Fig.~\ref{SkwNb25-64}(a), $S(k,\omega)$ is similar to the SF case in
Fig.~\ref{fullfillomega}(c): no gap is visible and the excitation
energies go to zero smoothly as $k\to 0$. Figure~\ref{SkwNb25-64}(b)
shows the same for $N_b=64$ where there is coexistence of MI and SF
phases. There is a marked difference in the behavior of $S(k,\omega)$
compared to the pure SF case: two branches of excitation peaks are
clearly visible in the form of two ridges. The low energy branch is
seen to be very similar to $S(k,\omega)$ in the SF case,
Fig.~\ref{SkwNb25-64}(a), saturating at about $\omega\approx 3t$, and
going smoothly to zero. The second family of peaks saturates at a
higher energy, $\omega \approx 8t$ and seems to make its first
appearance at about $k\approx 11$ in units of $2\pi/L$ with an $\omega
\approx 3t$.

\begin{figure}
\psfig{file=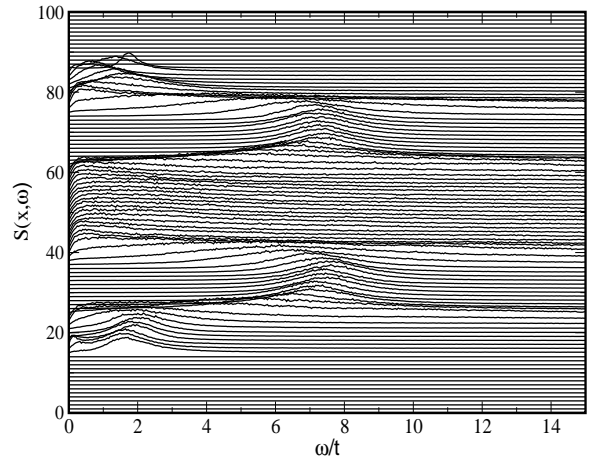,height=3in,width=3.5in,angle=-90}
\caption{$S(x,\omega)$ for the system in Fig.~\ref{SkwNb25-64}(b)
and shows the origin of gapped and ungapped excitation branches are
the SF and MI regions.}
\label{samesite}
\end{figure}

To interpret these two excitation branches, we examine $S(x,\omega)$,
the imaginary time Laplace transform of $\langle
n(x,0)n(x,\tau)\rangle$, the same site but time-separated
density-density correlation function. This is shown in
Fig.~\ref{samesite} which clearly identifies excitation branches with
position in the trap. $S(x,\omega)$ includes contributions from all
$k$ values. The very low energy {\it phonon} excitations arise from
the SF regions in the outer wings and the center of the system. The
middle excitations, $\omega \sim 2t-3t$ appear prominently in the
squeezed SF region between the MI and the outer edges of the
system. The value of $\omega$ at the peak corresponds well to the
saturation value, at high $k$, of the SF branch of $S(k,\omega)$ in
Fig.~\ref{SkwNb25-64}. This effect was observed in all our simulations
where a MI phase is present.  In these outer regions the superfluid is
squeezed between the MI and the rapidly increasing confining potential
which appears to enhance the spectral weight of the high $\omega$
excitations and make them so prominent.  Finally, the peaks in the two
MI regions in Fig.~\ref{samesite} agree well with the saturation value
of the higher $\omega$ branch in Fig.~\ref{SkwNb25-64}(b). This
confirms our interpretation that this branch corresponds to
excitations in the Mott phase. As further evidence, we mention that
the $S(x,\omega)$ for the uniform system in the MI phase at $U=4t$
(Fig.~\ref{fullfillomega}(a)) has its peak at $\omega\approx 8t$, like
the Mott regions in Fig.~\ref{SkwNb25-64}(b).

The interpretation of the higher $\omega$ excitation branch in
$S(k,\omega)$ as being due to the MI regions (Fig.~\ref{samesite}) and
the absence of a gap in Fig.~\ref{SkwNb25-64} show that even when MI
regions are present, the system is gapless! Figure~\ref{samesite}
shows that {\it local} excitation energies in the MI regions
correspond to those in MI phase of the uniform system at the same $U$.

Another very striking feature in Fig.~\ref{samesite} is the behavior
of $S(x,\omega)$ at the boundary between SF and MI sites: The peak in
$S(x,\omega)$ is very broad and extends from $\omega=0$ to the largest
values examined ($\omega=20t$) which indicates a {\it diverging
correlation length} in the time direction, $\xi_t(x)$. A diverging
$\xi_t$ means that $\langle n(x,\tau)n(x,0)\rangle$ decays as a power
law in $\tau$ and characterizes a quantum phase transition. In this
case, since it happens only locally, the divergence of $\xi_t(x)$ is
renewed evidence of the {\it local quantum criticality} already
discussed in~\cite{PRL02,rigol}.

\begin{figure}
\psfig{file=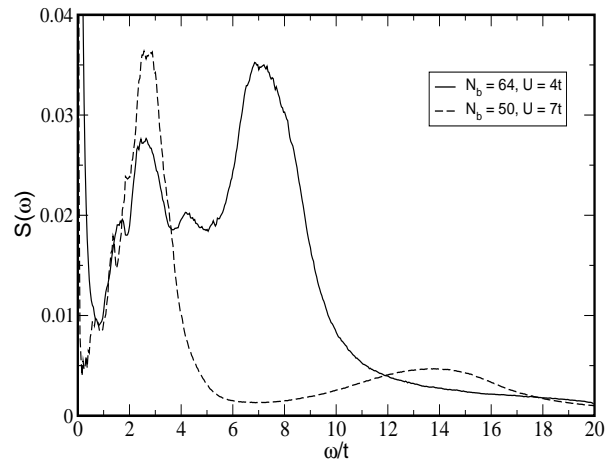,height=3in,width=3.5in,angle=-90}
\caption{$S(\omega)$ for $N_b=64, \, U=4t$ and for $N_b=50, \, U=7t$
  which has only a MI in the middle.}
\label{fwhm}
\end{figure}

To measure the total response of the system, summed over all momenta,
we use
\begin{equation}
S(\omega) \equiv \sum_k S(k,\omega).
\label{omegaprime}
\end{equation}
This quantity is shown in Fig.~\ref{fwhm} for the system shown in
Fig.~\ref{SkwNb25-64}(b) (same as Fig.~\ref{samesite}) and also for
$N_b=50, \, U=7t$ which has pure MI in the center surrounded by SF at
the edges. A two peak structure is seen where the first peak appears
not to depend on $U$ whereas the second peak does. The first peak,
therefore, is due to the SF component of the system and has an
$\omega$ which corresponds very well to the saturation value of the SF
branches of $S(k,\omega)$ in Figs.~\ref{SkwNb25-64}(a,b). On the other
hand, the higher peaks depend on the value of $U$ and both correspond
to the saturation values of the MI branches in $S(k,\omega)$ (see
Fig.~\ref{SkwNb25-64}(b)). Both peaks come from the high $k$ values of
the SF and MI branches because those values have the larger spectral
weights which therefore dominate when the sum over $k$ is
taken. Recall that the $S(k,\omega)$ in the figures have been rescaled
for better visibility and do not show the true spectral weights, but
Fig.~\ref{fwhm} does.

Our two-peak structure, Fig.~\ref{fwhm}, resembles the excitation
spectrum in Fig.2(a) in~\cite{StofEss04} for $^{87}$Rb on
one-dimensional lattices. Both quantities measure the total response
of the system to a given frequency, but they are not the same and the
connection between them is not clear. Although $S(k,\omega)$ does
detect the presence of MI regions, it also shows there is {\it no
overall gap} but that {\it local} excitation energies do correspond to
gap energies in the uniform system at the same $U$. It is possible
that the SF peak we detect is missed by reference~\cite{StofEss04}
because the amplitude with which they shake their system is very large
($20\%$ of the optical lattice potential).

We have presented the first Quantum Monte Carlo study of the dynamical
structure factor, $S(k,\omega)$, for the Hubbard model on uniform and
confined optical lattices.  The uniform system exhibits linear
dispersion (phonons) for the SF and a clear excitation gap in the MI
in agreement with the directly computed gap~\cite{GGBRTS90}. In the
absence of MI regions in the confined system, the $S(k,\omega)$
behavior is very similar to the uniform case.

However, when MI regions are present (Figs.~\ref{SkwNb25-64}(b) and
\ref{samesite}), we find that although $S(k,\omega)$ shows the
presence of such regions, it also clearly shows the {\it absence} of a
gap: the system is globally compressible and gapless~\cite{PRL02}. The
presence of the MI regions is deduced from the presence of two
excitation branches in $S(k,\omega)$ or from the double peak structure
of $S(\omega)$ (Fig.~\ref{fwhm}). We note, as discussed above, that
the second peak in Fig.~\ref{fwhm} corresponds to MI regions. In
addition, we note that $S(k,\omega)$ and $S(\omega)$ give information
about {\it local} excitations in the trap. For example, by comparing
$S(k,\omega)$ with $S(x,\omega)$ (Figs.~\ref{SkwNb25-64}(b) and
\ref{samesite}) we have found that the first peak in $S(\omega)$
represents excitations in the SF regions at the edges of the system
while the second peak represents the excitations in the MI
regions. The value of the local excitation energy is consistent with
the gap value for the uniform system at the same $U$. This is
consistent with the local quantum criticality view~\cite{PRL02,rigol}
for which we present new compelling evidence in Fig.~\ref{samesite}.
Our conclusions can be tested experimentally since the dynamic
structure factor $S(k,\omega)$ (and therefore, $S(\omega)$) can be
measured experimentally using Bragg
spectroscopy\cite{Braggspec,DSFBH}.

\noindent
\underbar{Acknowledgements} We acknowledge helpful discussions with
M. Troyer, H.T.C. Stoof and C. Checker. R.T.S. was supported by
NSF-DMR-0312261 and NSF-INT-0124863, PJHD is supported by Stichting
FOM.

\end{document}